\documentclass[aps,amsfonts,nofootinbib,superscriptaddress,preprint]{revtex4-1}
\usepackage{comment}
\usepackage[normalem]{ulem} 
\makeatletter
\usepackage{amsmath}
\usepackage{amssymb}
\usepackage{appendix}
\usepackage{amsbsy}
\usepackage{bm}
\usepackage{epsfig}
\usepackage{color}
\usepackage{slashed}
\usepackage{soul}
\usepackage{subcaption}

\usepackage[T1]{fontenc}
\usepackage{cancel}
\newcommand{\ee}{\end{equation}}
\newcommand{\bb}{\begin{equation}}
\newcommand{\eqb}{\begin{eqnarray}}
\newcommand{\eqf}{\end{eqnarray}}

\newcommand{\1}{{\'{\i}}}

\def\1{\'{\i}}

\def\1{\'{\i}}

\begin{document}
\title{ Kinetic and Magnetic Mixing with Antisymmetric Gauge Fields }
\author{Jorge Gamboa}
\email{jorge.gamboa@usach.cl}
\affiliation{Departmento de F\1sica, Universidad de Santiago de Chile, Casilla 307, Santiago, Chile}
\author{Fernando M\'endez}
\email{fernando.mendez@usach.cl}
\affiliation{Departmento de F\1sica, Universidad de Santiago de Chile, Casilla 307, Santiago, Chile}
\author{Justo L\'opez-Sarri\'on}
\email{jujlopezsa@unal.edu.co}
\affiliation{Departamento de Fisica Teorica, Atomica y Optica, Universidad de Valladolid,
47011 Valladolid, Spain 
\\ 
Departamento de F\1sica, Universidad Nacional de Colombia,
111321 Bogota, Colombia}
\begin{abstract}
A general procedure to describe the coupling $U_A (1) \times U_B (1)$ between antisymmetric gauge fields is proposed. For vector gauge theories the inclusion of magnetic mixing in the hidden sector induces millicharges -in principle- observable. We extend the analysis to antisymmetric fields and the extension to higher order monopoles is discussed. A modification of the model discussed in \cite{Ibarra} with massless antisymmetric fields as dark matter is also considered and the total cross section ratio are found and discussed. \end{abstract}
\maketitle
\section{Introduction }

Antisymmetric fields in particle physics are mainly used to describe resonances or the dynamics of particles with massive spin-$1$ at low energies such as the $\rho$-meson or the $a_1(1260)$ resonance (see also \cite{Hayashi}). 

However, the antisymmetric fields do not appear as fundamental descriptions of the elementary particles physics spectrum. The situation is different in string or relativistic membrane theory where antisymmetric fields are the natural way to incorporate interactions at the action level \cite{ramond}. 

An exception to  previous  arguments   was proposed in  \cite{Ibarra}
where an antisymmetric field has been  considered. It turns out to be stable, and transforms as a singlet under the  standard model gauge group. It is also massive, and for light masses, it could be a natural candidate for dark matter.

The idea of including antisymmetric fields, on the other hand,  is interesting by  itself. Indeed, these  were devised as a way to write fundamental fields disguised differently but retaining the same physics \cite{deser}, and therefore, the  idea of considering antisymmetric fields as another portal for dark matter is an exciting  possibility. 

In this note, we would like to explore a different scenario, namely, the existence of an antisymmetric hidden field, an U(1) gauge field, with the possibility to interact with the visible sector through, for example, magnetic and kinetic mixing.

In concrete, in the present paper, we study the problem of considering antisymmetric fields as a new portal to analyze dark matter and its topological properties. We will start by considering the case of vector fields as a warm-up exercise, and we 
will re-derive some previously known results  \cite{jae} from a  different point of view, and then, in section III, we will
consider antisymmetric fields, and the construction of kinetic mixing.  
In section IV we provide an example of these ideas and we analyze  possible phenomenological implications. Finally, in section V, the conclusions and the scope of our results are explained.

\section{Abelian $U (1) \times U(1)$ theories}

We start this section considering two $U(1)$ gauge fields which are coupled according to 
\bb
{\cal L} = -\frac{1}{4} F_{\mu \nu}(A) F^{\mu \nu} (A) -\frac{1}{4} F_{\mu \nu}(B) F^{\mu \nu} (B)  + A_\mu J^\mu (B),  \label{2}
\ee
  where $A_\mu$ and $B_\mu$  (visible and hidden sectors, respectively) transform under $U(1)\times U(1)$ as
  \eqb
  A'_\mu &=& A_\mu +     \partial_\mu \Lambda, \nonumber
  \\
 B'_\mu &=& B_\mu +     \partial_\mu {\bar \Lambda}, \label{4}
 \eqf
where $\Lambda,\bar{\Lambda}$ are the two independent parameters of the gauge transformations.

 Since $A_\mu$ must transform as in  (\ref{4})   {under  the two $U(1)$ gauge groups,  the invariance of the action
 implies the conservation of  $J^\mu (B)$, that is} 
\bb
 \partial_\mu J^\mu (B) =0. \label{5}
\ee

The external currents $j^\mu,$ (matter currents)  are added through the replacement
\bb
J^\mu (B) \to J^\mu (B) + j^\mu= J^\mu_{\mbox{\tiny total}}.
\ee
For instance, in QED, such a current turn out to be ${\bar \psi} \gamma^\mu \psi$ or, in scalar electrodynamics, this current corresponds to  $\phi^* {    \partial_\mu} \phi + \mbox{h.c}$. The total current is also conserved, since matter current is conserved due to the gauge invariance.
%
%
%

 In the gauge sector,   the conservation equation (\ref{5}) {,  together with the invariance of the action under} $U(1)$ { 
 in sector} $B$,  {allows writing } the  most general solution for $J^\mu (B)$ as \cite{cortes}
 \bb
  J^\mu (B) = c_1     \partial_\nu F^{\mu \nu} (B) + c_2      \partial_\nu {\tilde F}^{\mu \nu} (B), \label{6}
  \ee
 where   {the dual  $\tilde{F}^{\mu\nu} = \frac12\epsilon^{\mu\nu\lambda\rho}F_{\lambda\rho}$. The 
 constant coefficients $c_1$ and $c_2$ can be fixed, for example,}
  by comparing  {with} other known results,  {as we will show at the end of the present section.}

 Replacing (\ref{6}) in (\ref{2}) and integrating by parts, we obtain
 \bb
 {\cal L} = -\frac{1}{4} F_{\mu \nu}(A) F^{\mu \nu} (A) -\frac{1}{4} F_{\mu \nu}(B) F^{\mu \nu} (B)  +
 \frac{c_1}{ {2}} F_{\mu \nu} (A)F^{\mu \nu} (B) + \frac{c_2}{ {2}}
 F_{\mu \nu} (A){\tilde F}^{\mu \nu} (B),
 \label{7}
\ee
 {which} reproduces the kinetic mixing \cite{holdom}  {term} and also the magnetic  {contribution}
 which are usually introduced  {on gauge invariant grounds.}  The procedure  {implemented} 
 here states how to couple two gauge fields $U (1)$ using  similar arguments  {and} the Noether theorem.

  {Given the Lagrangian (\ref{2}), the field equations turn out to be}
 \eqb
     \partial_\mu F^{\mu \nu} (A) &=& c_1     \partial_\mu F^{\nu \mu} (B) + c_2     \partial_\mu {\tilde F}^{\nu \mu} (B), \label{8}
 \\
     \partial_\mu F^{\mu \nu} (B) &=& c_1     \partial_\mu F^{\nu \mu} (A) + c_2     \partial_\mu {\tilde F}^{\nu \mu} (A).  \label{9}
 \eqf

  { As stated above, the gauge field $B_\mu$ belongs to a hidden sector, while $A_\mu$ describes photons
 in the visible one.  Since there is no evidence of  Dirac magnetic} monopoles in the visible sector,   we impose $    \partial_\mu {\tilde F}^{\nu \mu} (A) =0$,  so that  (\ref{8}) and (\ref{9}) become 
 \eqb
     \partial_\mu F^{\mu \nu} (A) &=& c_1     \partial_\mu F^{\nu \mu} (B) + c_2     \partial_\mu {\tilde F}^{\nu \mu} (B), \nonumber
 \\
     \partial_\mu F^{\mu \nu} (B) &=& c_1     \partial_\mu F^{\nu \mu} (A), 
  \label{9cc}
 \eqf 
 and therefore
 \bb
     \partial_\mu F^{\mu \nu} (A) = \frac{c_2}{c^2_1-1}     \partial_\mu {\tilde F}^{\mu \nu} (B). \label{10x}
 \ee  
 
  {Let us stress that the visible sector has no topological obstruction in the present formulation, but  such 
 restriction does not need to be imposed in the hidden one. Indeed, equation} (\ref{10x})  {corresponds to} the  Maxwell equations in the presence of an  external source,  {indicating that hidden magnetic} monopoles must be taken into 
 account \cite{yang,wy,goddard},  {except for the case $c_1=\pm1$, when the set  (\ref{9cc}) is equivalent to 
\begin{equation}
    \partial_\mu \big(F^{\mu\nu}(A)\pm F^{\mu\nu}(B)\big) = 0, \quad     \partial_\mu\tilde{F}^{\mu\nu}(B) =0.
\end{equation}
}
 
  { Let us analyze the case $c_1 \neq \pm 1$, for which the hidden magnetic monopoles act as a source
 for the visible sector.  We consider the static case, and then  the Gauss law reads}
 \bb
 \nabla \cdot{\bf E}_A = \frac{c_2}{c^2_1-1}\nabla\cdot{\bf B}_B. \label{10xx}
 \ee 
  { For the static, hidden magnetic monopole, we choose the Dirac solution} $\nabla\cdot{\bf B}_B =
   {4\pi g_{\mbox{\tiny{B}}}}\,\delta ({\bf x})$, with  $g_{\mbox{\tiny{B}}}$, the hidden magnetic charge.  {We can interpret the r.h.s. in (\ref{10xx})
  as the electric charge density which is the source of the visible electric field, that is}
 \bb
\nabla \cdot{\bf E}_A = 4\pi \rho_A= 4\pi g_{\mbox{\tiny{B}}}\frac{c_2}{c^2_1-1}\,\delta ({\bf x}),  \label{vi1}
 \ee
 implying the effective visible electric charge
 \bb
 \rho_A= \frac{2 n \pi}{q_{\mbox{\tiny{B}}}}\frac{c_2}{c^2_1-1}\,\delta ({\bf x}),
  \label{vi01}
 \ee
 where the Dirac quantization condition  {in the hidden sector} has been used.  {Note that we are 
 also assuming the existence of electrically charged particles in the hidden sector, with charges $q_{\mbox{\tiny{B}}}.$}

The coefficients $c_1$ and $c_2$ can be  {identified by comparison  with similar terms discussed in the literature;} by comparing  {with the work by Holdom \cite{holdom}, where two 
$U(1)$ gauge groups where considered, }  $c_1$ is  {minus} the kinetic mixing parameter -- originally  denoted  $\chi$ 
in \cite{holdom} --  namely
 \bb
 c_1= - \chi.
 \label{ki1}
 \ee

 {Coefficient  $c_2$, which  is more subtle, can be identified with similar terms in the Lagrangian  
considered by Br\"ummer,   Jaeckel,  and Khoze in}  \cite{jae} where  { the effects of $\theta$-terms  \cite{witten} 
mixing  field strengths in theories with an extra $U(1)$ (hidden) gauge group, was considered.}
In the present work, for the magnetic kinetic mixing term 
 $c_2\, F_{\mu \nu} (A) {\tilde F}^{\mu \nu} (B) \propto  c_2\, {\bf E}_A \cdot{\bf B}_B$, we can assume that ${\bf B}_B$ contains both  a regular magnetic field and a point magnetic monopole, that is
 \bb
 {\bf B}_B = {\bf B}_{B}^{d} +\frac{g_{\mbox{\tiny{B}}}}{4\pi} \frac{{\bf r}}{|{\bf r}|^3}, 
 \ee
where ${\bf B}_{B}^{d}$ is the dynamical hidden magnetic field,  {and the second term is the static hidden magnetic
monopole with magnetic charge $g_{\mbox{\tiny{B}}}$. Considering as before a static electric field  ${\bf E} _A$,}
we find    {the relation between the $c_2$ coefficient and the $\theta$-term}
   \bb
  c_2 = -\frac{\theta }{\pi}.  \label{theta}
  \ee
  
Finally, we find the effective electric charge $q_A^{\mbox{\tiny{eff}}}$ due to a source of hidden photons with a magnetic mixing.
  \bb
  q_A^{\mbox{\tiny{eff}}} = \frac{2n\theta}{q_{\mbox{\tiny{B}}}}\frac{1}{1-\chi^2},
  \ee

 \section{antisymmetric tensor fields kinetic and magnetic mixing}

 The above idea can be directly generalized by considering instead of potential $A_\mu$, antisymmetric tensors $A_{\mu_1 \mu_2 \cdots \mu_{p}}$. As far as we know,  {the} first discussion in this direction is due to Kalb and Ramond \cite{ramond} {,} who introduced a second order antisymmetric tensor in order to incorporate new couplings in string theory.

 The basic idea underlying the Kalb-Ramond construction is invariance under reparametrizations in the world-sheet and the extension to higher-dimensional extended objects.  Although classically it can be carried out, it has intricate technical,
and topological subtleties which began to be studied in \cite{teitelboim} and this topic continues to be an area of intense research \cite{heck}. 
 
The idea that we will develop in this section is similar to the construction of Kalb and Ramond \cite{ramond} and Teitelboim \cite{teitelboim} (see also \cite{orland}), but it is instead the invariance under diffeomorphism we will have the symmetry $U (1) \times U (1)$ and  the action is 
 \bb
 S= \int d^Dx \, {\cal L}, \label{act}
 \ee 
 with  ${\cal L}$ given by 
 \bb
{\cal L} = -\frac{1}{2}\frac{1}{( {p+1})!} F^2_{\mu_1 \mu_2 \dots  \mu_{p+1}} (A) 
- \frac{1}{2}\frac{1}{( {p+1})!} F^2_{\mu_1 \mu_2 \dots  \mu_p} (B)  
+
 {\frac{1}{p!}} A_{\mu_1 \mu_2 \cdots \mu_{p}} J^{\mu_1 \mu_2 \cdots \mu_{p}} (B), 
 \label{13}
\ee
 where $A_{\mu_1 \mu_2 \cdots \mu_{p}}$ is an antisymmetric tensor and the \lq \lq strength\rq \rq $\,$  is
 \bb
 F_{\mu_1 \mu_2 \dots  \mu_{p+1}} =     \partial_{\mu_1}  A_{\mu_2 \mu_3 \cdots \mu_{ {p+1}}} -
     \partial_{\mu_2}  A_{\mu_1 \mu_3 \cdots \mu_{ {p+1}}} \cdots 
 -    \partial_{\mu_{ {p+1}}} A_{\mu_1\mu_2 \cdots\mu_p}, \label{14}
 \ee
 {and the usual definition $F^2=F_{\mu_1,\mu_2,\cdots\mu_{p+1}}F^{\mu_1,\mu_2,\cdots\mu_{p+1}}$.}

 Noticing that in (\ref{act}) the spacetime dimension is $D$,  { we choose} \cite{teitelboim}
 \[ 
 D= 2 \left(p+1\right),
 \]   
 {and then,} the dual tensor  is 
   \bb
 {\tilde F}^{\mu_1 \mu_2 \cdots \mu_{p+1}} = \frac{1}{p!} \epsilon^{\mu_1 \mu_2 \cdots \mu_{p+1} \nu_1 \nu_2 \cdots \nu_{p+1}} F_{\nu_1 \nu_2 \cdots \nu_{p+1}}. 
 \ee  
  
  The generalization of  {gauge transformations in} (\ref{4}), reads
  \eqb
  A'_{\mu_1 \mu_2 \cdots \mu_{p}} &=& A_{\mu_1 \mu_2 \cdots \mu_{p}} +     \partial_{[\mu_1}
  {{\Lambda}}_{\mu_2 \mu_3 \cdots \mu_{p}]}, \nonumber
  \\
 B'_{\mu_1 \mu_2 \cdots \mu_{p}} &=& B_{\mu_1 \mu_2 \cdots \mu_{p}} +     \partial_{[\mu_1}
  {\bar{\Lambda}}_{\mu_2 \mu_3 \cdots \mu_{p}]},
  \label{16}
 \eqf
  {with $\Lambda_{\mu_1,\mu_2,\dots,\mu_{p-1}}$ and  $\bar{\Lambda}_{\mu_1,\mu_2,\dots,\mu_{p-1}}$, two
 arbitrary antisymmetric tensors and the notation $[\dots]$ stands for fully antisymmetric index.}
 
 The  {field equations derived from Lagrangian (\ref{14}) are}
 \eqb
     \partial_{\mu_1} F^{\mu_1 \mu_2 \cdots \mu_{p+1}} (A) &=& J^{\mu_2 \mu_3 \cdots \mu_{ {p+1}}} (B), \label{cu1}
 \\
      \partial_{\mu_1} F^{\mu_1 \mu_2 \cdots \mu_{p+1}} (B) &=& J^{\mu_2 \mu_3 \cdots \mu_{ {p+1}}} (A). \label{cu2}
  \eqf

 The most general choice of $J^{ {\mu_1 \mu_2} \cdots \mu_{p}}$,  
 consistent with the conserving current  {condition,} 
 {\it i.e.} $    \partial_{ {\mu_1}} J^{\mu_1 \mu_2 \cdots \mu_{p}}=0$ is 
 \bb
 J^{\mu_1 \mu_2 \cdots \mu_{p}}  = c_1\,     \partial_{ {\nu}} F^{ {\nu \mu_1 \cdots \mu_{p}}} + c_2\,    \partial_{ {\nu}} {\tilde F}^{ {\nu \mu_1 \cdots \mu_{p}}}, 
 \label{curr1} 
 \ee
where $c_1$ and $c_2$ are determined below  {and the currents are evaluated in both sectors, $A$ and $B$.}

Assuming as above that there are no monopoles in the visible sector, that is $$    \partial_{\mu_1} {\tilde F}^{\mu_1 \mu_2 \cdots \mu_{p+1}}(A)=0, $$ 
the equations (\ref{cu1}) and (\ref{cu2}) are simplified to 
 \eqb
     \partial_{\mu_1} F^{\mu_1 \mu_2 \cdots \mu_{p+1}} (A) &=& c_1     \partial_{\mu_1} F^{\mu_1 \mu_2 \cdots \mu_{p+1}}  (B)+ c_2    \partial_{\mu_1} {\tilde F}^{\mu_1 \mu_2 \cdots \mu_{p+1}}(B), \nonumber
 \\
      \partial_{\mu_1} F^{\mu_1 \mu_2 \cdots \mu_{p+1}} (B) &=& c_1     \partial_{\mu_1} F^{\mu_1 \mu_2 \cdots \mu_{p+1}} (A). \label{cu4}
  \eqf
  
  Replacing the second equation in the first one, we find 
  \bb  
       \partial_{\mu_1} F^{\mu_1 \mu_2 \cdots \mu_{p+1}} (A) =  \frac{c_2}{1-c_1^2}     \partial_{\mu_1} {\tilde F}^{\mu_1 \mu_2 \cdots \mu_{p+1}}(B). \label{cu5}
 \ee  
       
Equation (\ref{cu5}) is the counterpart of (\ref{10x}).  {On the other hand, the condition}
 $    \partial_{\mu_1} {\tilde F}^{\mu_1 \mu_2 \cdots \mu_{p+1}}(B)=0$  is the analog of 
   the no monopole condition $\nabla\cdot{\bf B}=0$ in electrodynamics.  {Indeed, for } $    \partial_{\mu_1}{\tilde F}^{\mu_1 \mu_2 \cdots \mu_{p+1}}(B) \neq 0$,  { one should}  have higher rank monopoles, as was discussed in \cite{nepo,teitelboim2}.  
   
   
   However,   { to determine the presence of higher-rank monopoles  is cumbersome in this scheme, that is, to find a solution for}
   \bb
        \partial_{\mu_1}{\tilde F}^{\mu_1 \mu_2 \cdots \mu_{p+1}}(B) = j^{\mu_2 \cdots \mu_{p+1}} (B), 
    \nonumber
    \ee  
and it is better to proceed in analogy with the usual Dirac monopole and the Wu-Yang method \cite{wy}.  
   
    Noteworthy  that the current $ j $ appears as a higher rank-monopole source for the hidden sector but by (\ref{cu5}) it is also a source for the visible sector in full analogy with discussion II.
       
   If we adopt the notation $B^N$ and $B^S$ for the hidden potentials at the northern and southern poles, the difference $B^N-B^S$ must give the hidden magnetic flux
   \bb
   \int_{{\cal M}_p} ( B^N -B^S) =\int_{    \partial {\cal M}_p} d\Lambda = 4 \pi g_B (\#), \label{intw}
   \ee 
   where $(\#)$ is a notation for the linking numbers of the strings and the surface ${\cal M} _ {p + 1}$ which is a topological invariant \cite{henneaux}.
   
  The determination of the coefficients $c_1$ and $c_2$ is obtained following the same arguments of section II. 
   The coefficient $c_1$ is just the mixing parameter while $c_2$ corresponds to the \lq \lq vacuum angle\rq \rq.  
    
    \section{An Application}
    
    As an application of the ideas discussed above let us consider the following extension of the standard model
    \bb
    {\cal L} = {\cal L}_{\mbox{\tiny{ {SM}}}} +{\cal L}_1 + {\cal L}_{AB} + {\cal L}_{int}, \label{mod1}
    \ee
    where  ${\cal L}_{\mbox{\tiny{ {SM}}}}$ is the standard model Lagrangian,  {and}
    \eqb
    {\cal L}_1 &=&  -\frac{1}{2}\frac{1}{( {p+1})!} F^2 (A)   
    -\frac{1}{2}\frac{1}{( {p+1})!}  F^2(B), \label{mod2}
    \\
    {\cal L}_{AB}  &=&  A_{\mu_1 \mu_2 \cdots \mu_{p}} J^{\mu_1 \mu_2 \cdots \mu_{p}} (B), \label{mod3}   
        \eqf
    with $J^{\mu_1 \mu_2 \cdots \mu_{p}} (B)$ defined in (\ref{curr1}). 
    
   {In order to discuss a possible phenomenology, one should define a dimensional reduction scheme.
  For example,} {  {  if (\ref{curr1}) comes from the low energy limit of string theory,  the
   four-dimensional compactification forces the $p$-forms  {$A$ and $B$} to be $2$-forms of Kalb-Ramond 
   and redefinitions of the energy scales  {whose solely effect is a redefinition of } the parameters of the theory.   }}

 {  { Taking into account  {this dimensional reduction,} the interaction Lagrangian  ${\cal L}_{int}$  becomes} 
    \bb
    {\cal L}_{int} = \left(g_A  A_{\mu_1\mu_2} A^{\mu_1\mu_2} + g_B B_{\mu_1\mu_2} B^{\mu_1\mu_2} + g_{AB} A_{\mu_1\mu_2}B^{\mu_1\mu_2}\right) \, h^{\dagger} h, \label{mod4}
    \ee
    where $h$ is a charged scalar field (Higgs).
        
These couplings are all renormalizable.  {The interactions} correspond
 to the annihilation of  {the} antisymmetric tensors  {which are} additional fields of the standard model.    
The processes are depicted in figure Fig. 1.

\begin{figure}[h!]
\centering
\begin{subfigure}[t]{0.46 \textwidth}
\includegraphics[width=0.46\textwidth]{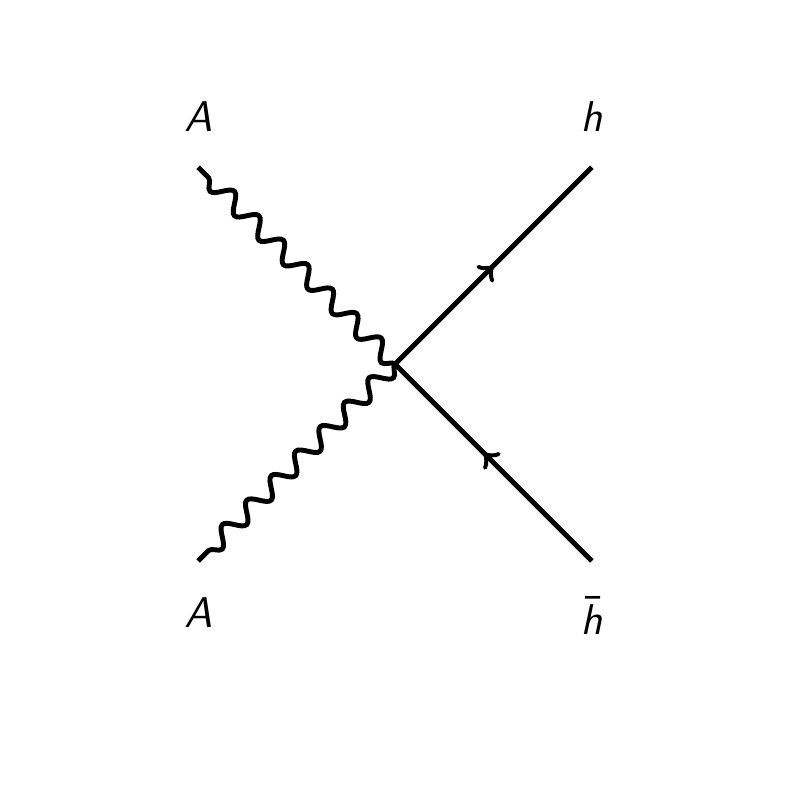}
\caption{}
\label{cone}
\end{subfigure}
\centering
\begin{subfigure}[t]{0.46 \textwidth}
\includegraphics[width=0.46\textwidth]{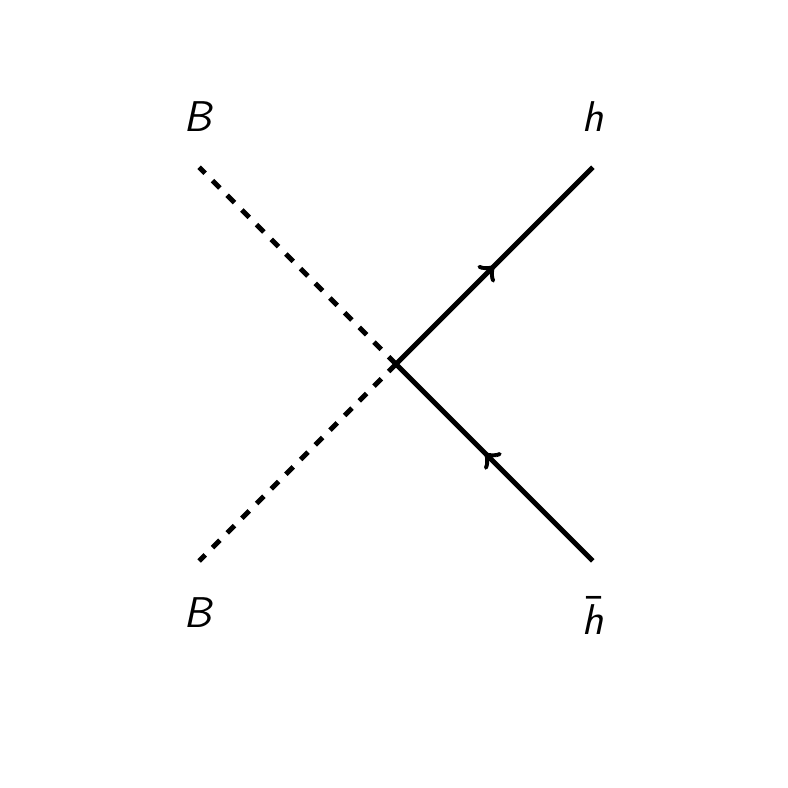}
\caption{}
\label{diag1}
\end{subfigure}
\centering
\begin{subfigure}[t]{0.46 \textwidth}
\includegraphics[width=0.46\textwidth]{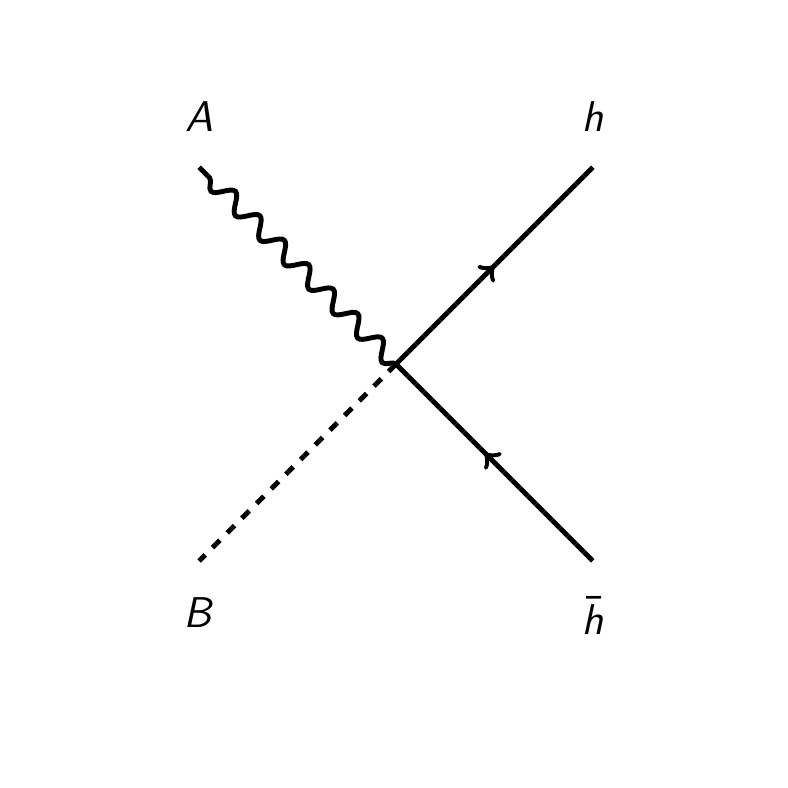}
\caption{}
\end{subfigure}
\centering
\caption{Three different processes from interacting Lagrangian (\ref{mod4}). Panel a) exhibits   the scattering  $A + A \to h^\dagger + h $,  {analogous to } the scattering    $\gamma \gamma \to e^+ e$. Panel b)  corresponds to the process $B + B \to h^\dagger h$ while panel c) shows  $A + B \to h^\dagger h$
scattering.}
\label{fig1}
\end{figure}

%
%

%


 {For example, } Fig. (\ref{cone}) describes $A+A \to h^\dagger +h$.  { Assuming large values of $s$,
 { the center of mass energy,}  the cross section can be calculated in  analogy to a Breit-Wheeler process in QED.}

The total cross section  { for the processes in figure (\ref{fig1}), under the same assumptions,}  are 
\eqb
\sigma(AA\to h^\dagger h)&=& \frac{g_A m_h}{s} \ln \left(\frac{\sqrt{s}}{m_h}\right) \nonumber 
\\
\sigma(BB\to h^\dagger h)&=& \frac{g_B m_h}{s} \ln \left(\frac{\sqrt{s}}{m_h}\right), \nonumber 
\\
 \sigma(AB\to h^\dagger h)&=& \frac{g_{AB} m_h}{s} \ln \left(\frac{\sqrt{s}}{m_h}\right),   \nonumber 
\eqf
implying the following cross sections ratios 
\eqb
\frac{\sigma(AA\to h^\dagger h)}{\sigma(BB\to h^\dagger h)} &=& \frac{g_A}{g_B}, \nonumber 
\\
\frac{\sigma(AA\to h^\dagger h)}{\sigma(AB\to h^\dagger h)} &=&\frac{g_A}{g_{AB}}, \label{ratio} 
\\
\frac{\sigma(BB\to h^\dagger h)}{\sigma(AB\to h^\dagger h)} &=&\frac{g_B}{g_{AB}}. \nonumber
\eqf

Ratios (\ref{ratio}) are simpler and more accessible in a model like the one we have discussed here.
       
      \section{Conclusions and outlook}
   The description of gauge theories with kinetic and magnetic mixing is an approach that has been intensively investigated in recent years as a way of describing dark matter. If magnetic mixing is included in the hidden sector, the possibility of seeing a millicharge effect would be possible.  
   
   In order to estimate the effects of millicharges one can proceed as follows: in section II we have seen 
   that the presence of magnetic monopoles in the hidden sector induces the visible charge density (\ref{vi01}) 
   which can be interpreted as the millicharge (n=1)    
   \bb 
   q_{\mbox{\tiny{milli}}} = \frac{2\theta}{q_{\mbox{\tiny{B}}}}\frac{1}{1-\chi^2} \approx \frac{2\theta}{q_{\mbox{\tiny{B}}}}, 
   \ee
   which is a contribution due entirely to magnetic mixing. 
   
      The force produced by a millicharge compared to the Coulomb force between two electrons is    
      \[
      \left|\frac{F_{\mbox{\tiny{milli}}}}{F_{\mbox{\tiny{Coulomb}}}} \right| \approx 
      \left(\frac{2\theta}{eq_{\mbox{\tiny{B}}}}\right)^2, 
      \]
     while the force between an electron and a millicharge compared to the Coulomb
     force between two electrons is 
       \[
      \left|\frac{F_{\mbox{\tiny{milli}}-e }}{F_{\mbox{\tiny{e-e}}}} \right| \approx 
      \frac{2\theta}{eq_{\mbox{\tiny{B}}}}, 
      \]
and the effects of the millicharges cannot be neglected.  
      
      The estimation of $\theta$ is central to cold dark matter phenomenology because the measurement of $\theta$ is an indirect measure that can be associated with axion detection via the Peccei-Quinn mechanism. Thus, the problem of estimating 
      $\theta$ is moved to exploring the values of $g/M^2$ in axion phenomenology \cite{axionrev}.
      
In this paper we have proposed an extension of  the kinetic mixing idea to antisymmetric fields which could have implications in the search for physics beyond the standard model. Indeed, we have shown that this 
procedure gives rise to new   decay channels. From here it is possible, in principle,  to extract bounds for the coupling constants. Thus, antisymmetric fields can also be seen as another way of describing fundamental fields as was discussed long ago by Deser and Townsend (see \cite{deser}).
 
        We would like to thank Prof. Fidel A. Schaposnik for  interesting discussions in the initial stages of this work. This research was partially supported by Dicyt 042131GR (J.G.) and 041931MF (F.M.) and  by Fundacion ONCE with grant Oportunidad al Talento (J.L.S.).

\end{document}